\documentclass[aps,prl,twocolumn,noshowpacs,nofootinbib,superscriptaddress,groupedaddress]{revtex4}
\usepackage{graphicx}  
\usepackage{dcolumn}   
\usepackage{bm}        
\usepackage{amssymb}   

\usepackage[utf8]{inputenc}

\usepackage{amsmath}

\usepackage{amsfonts}

\usepackage{textcomp}

\usepackage{bm}

\usepackage{color}

\usepackage{verbatim}

\usepackage{subfig}

\begin{document}

\title{Hagedorn temperature and physics of black holes}

\author{Thomas G. Mertens}
\email{tmertens@princeton.edu}
\affiliation{Joseph Henry Laboratories, Princeton University, Princeton, NJ 08544, USA,\\
Ghent University, Department of Physics and Astronomy
Krijgslaan, 281-S9, 9000 Gent, Belgium}
\author{Henri Verschelde}
\email{henri.verschelde@ugent.be}
\affiliation{ 
Ghent University, Department of Physics and Astronomy
Krijgslaan, 281-S9, 9000 Gent, Belgium}
\author{Valentin I. Zakharov}
\email{vzakharov@itep.ru}
\affiliation{ITEP, B. Cheremushkinskaya 25, Moscow, 117218 Russia,\\
Max-Planck Institut f\"ur Physik, F\"ohringer Ring 6, M\"unchen, Germany,\\
School  of  Biomedicine,  Far  Eastern  Federal  University,  Sukhanova  str  8, 
 Vladivostok  690950
Russia}

\begin{abstract}
A mini-review devoted to some implications of the Hagedorn temperature
for black hole physics.
The existence of a limiting temperature is a generic feature of string models.
The Hagedorn temperature was introduced first in the context of hadronic physics.
Nowadays, the emphasis is shifted to fundamental strings
which might be a necessary ingredient to obtain a consistent theory
of black holes. The point is that, in field theory, the local temperature
close to the horizon could be arbitrarily high, and this
observation is difficult to reconcile with the finiteness of the entropy of black holes.
After preliminary remarks, we review our
recent attempt to evaluate 
the entropy of large black holes in terms of fundamental strings. 
We also speculate on implications for dynamics of large-$N_c$ gauge
theories arising within holographic models.
  
\end{abstract}

\maketitle

\section{Introduction.}

\subsection{From hadrons to black holes}

This talk was presented at a session devoted to the 50th anniversary of the 
introduction of the Hagedorn temperature \cite{hagedorn}. Originally, the
Hagedorn temperature $T_H$ was discussed in connection with hadronic   
 physics. Alternatively, one can say that it referred to a mass scale of the order
of the pion mass, $T_H~\sim~m_{\pi}$. We will discuss the physics of black holes, 
or, more precisely, the properties of the black-hole horizon.
In field theory, the local temperature close to the horizon can be arbitrarily
high, and this is known to be inconsistent with finiteness of the
Bekenstein-Hawking entropy of black holes, see, in particular,
\cite{thooft}\cite{'tHooft:1984re}. The inconsistency becomes manifest at the gravitational scale, 
or at $T\sim M_{\text{Planck}}$. This observation serves as a motivation \cite{susskind}
to introduce strings at this scale. It is within this framework that we make our remarks 
on the explicit evaluation of the Bekenstein-Hawking entropy, see \cite{mertens} and references
therein. Nowadays, the literature on the subject of strings and black holes is huge.
Because of the format of the talk we limit our list of references to only a few papers.
General background can be found in \cite{susskind2}. 

It should be remarked that the recent firewall paradox (supposedly burning up infalling observers) has rekindled the interest in horizon physics (see \cite{Almheiri:2012rt} and subsequent work). We however are interested in the physics as described by static observers and are hence (a priori) safely away from speculating on the experience of infalling observers. 

The outline of the talk is as follows. In the Introduction we describe briefly
how the notion of a limiting temperature arises within a generic string picture.  
In Section 2 we remind the reader of the basics of the black holes. 
In Section 3 we address the issue of evaluating black-hole entropy within string theory.
In Section 4 we discuss briefly possible phenomenological implications
within holographic models.
 \subsection{Hagedorn temperature}

As a starting point, we can choose the assumption made by R. Hagedorn
\cite{hagedorn}
that the density of hadronic states $\omega(E)$ at large energy $E$
grows exponentially:
\begin{equation}\label{spectrum}
\omega(E)~\sim~\exp\big(\beta_HE\big)~~, \text{where}~~~\beta_H~\sim~m_{\pi}^{-1}.
\end{equation}
Then the partition function
\begin{equation}\label{divergence}
z~=~\int_0^{\infty}dE\omega(E)e^{-\beta E}
\end{equation}
exists only as far as $\beta~>~\beta_H$.
In the microcanonical language, there is a ``limiting temperature" $T_H$:
\begin{equation}
\frac{1}{T}~=~\frac{\partial S}{\partial E}
\end{equation}
and \begin{equation}
T~<~T_H~,~~~~T_H~\equiv~\frac{1}{\beta_H}~=~ \text{const}~~.
\end{equation}
The physics behind this remarkable phenomenon is
actually quite simple: if we pump energy into the system, new higher-mass states are 
produced rather than that the energy of already existing states is increased. The increasing of energy of existing states 
would mean an increasing temperature. 
The dominance of production of new massive states
manifests the emergence of a limiting temperature.

\subsection{Hagedorn temperature and strings}

In view of the fact that the spectrum (\ref{spectrum}) leads to such a drastic conclusion
as the existence of a limiting temperature in nature, 
we should probably re-examine the reasons for introducing the exponential 
spectrum itself.  

The strongest support for the assumption (\ref{spectrum}) comes from the string model
of hadrons.
And, in turn, the strongest point of the string model is that it reproduces
linear Regge trajectories.
Indeed, the energy of a string of length $L$ is given by:
\begin{equation}\label{tension}
E_{\text{string}}~=~\sigma\cdot L,~~\sigma~\equiv~(2\pi\alpha')^{-1}~,
\end{equation}
where $\sigma$ is the string tension. 
For a rotating string
\begin{equation}\label{rotation}
M^2~=~\frac{1}{\alpha'}J,
\end{equation}
where $J$ is the total angular momentum. In other words, the 
Regge trajectories are linear. Moreover, 
the density of states is indeed exponential at high energy:
\begin{equation}\label{exponential}
\omega(E)~\sim~\frac{\exp (\beta_HE)}{E^{1+D/2}}~,
\end{equation}
where $\beta_H\sim \sqrt{\alpha'}$ and
$D$ is the number of (non-compact) spatial directions.
Derivations of Eqs (\ref{tension}), (\ref{rotation}), (\ref{exponential})
can be found in standard textbooks.

\subsection{Limiting temperature vs phase transition}

At first sight, the argumentation above looks strong enough and
we could expect the existence of a limiting temperature. In fact,
it was realized a long time ago that there is a viable alternative
to the introduction of the Hagedorn temperature. 
Namely, one can argue 
that there is a phase transition. While at low temperatures hadrons appear
to be fundamental they are in fact composite
and are built up by quarks and gluons. Thus, at some critical temperature
$T_{cr}$ there is a phase transition to deconfinement. This phase transition
has been observed and studied thermodynamically in great detail through lattice
simulations for various non-Abelian gauge groups,
including the realistic case
of quantum chromodynamics. 

The precise relation between $T_H$ and $T_{cr}$ remains somewhat obscure.
Phenomenologically, it is obvious that one should have $T_H> T_{cr}$,
where $T_H$ is defined within the string model. How close $T_H$ is to
$T_{cr}$, remains unclear because of the uncertainties of the string models
of hadrons.

 Thus, we can say that the existence of  $T_{cr}$ can be traced back to
the fact that
it is the field theory which is fundamental, not the hadronic strings model.

As we will see next, the quantum field theory (QFT) becomes, in turn, problematic
at the gravitational scale. This is revealed by considering black holes.

\section{Black holes. Preliminaries}
Let us first remind the reader a few well-known equations 
concerning black holes.
The
Schwarzschild geometry reads as
\begin{equation}
ds^2~=~-\Big(1-\frac{2G_NM}{r}\Big)dt^2~+~\Big(1-\frac{2G_NM}{r}\Big)^{-1}dr^2 +
d{\bf x}^2_{\perp},
\end{equation}
where $M$ is the mass of the central body, or black hole in our case and 
$G_N$ is the Newton
constant. Note that the $G_{00}$ component
of the metric vanishes at the horizon $r_H~=~2G_NM$.

The thermodynamic entropy is proportional to the area of the black hole:
\begin{equation}\label{bhentropy}
S_{BH}~=~\frac{\text{Area}}{4G_N}
\end{equation}
and there is Hawking radiation with temperature
\begin{equation}
\beta_{\text{Hawking}}~=~8\pi G_N M~~.
\end{equation} 
Close to the horizon, it is useful to introduce the
distance $\rho$ to the horizon,
$\rho~=~\sqrt{8G_NM(r-2G_NM)}$.
Then for $\rho\ll 4G_NM$  
\begin{equation}
ds^2_{\text{Rindler}}~=~-\frac{\rho^2}{(4G_NM)^2}dt^2+d\rho^2+d{\bf x}^2_{\perp}.\nonumber
\end{equation}
Many results apply just in this limit of the so-called Rindler space.
For Euclidean time $\tau$,
\begin{equation}
ds^2_{\text{Euclidean}}~=~\frac{\rho^2}{(4G_NM)^2}d\tau^2+d\rho^2+d{\bf x}^2_{\perp}, \nonumber
\end{equation}
which is flat space in polar coordinates.
Moreover, the $\tau$-coordinate is  periodic,
\begin{equation}
\tau~\sim~\tau~+~\beta_{\text{Rindler}}~~,
\end{equation}
with 
$\beta_{\text{Rindler}}~=~8\pi G _NM~\equiv~\beta_{\text{Hawking}}$.

One can say that a 
black hole provides a ``lab" to study temperatures arbitrarily high. Indeed,
near the horizon  the
blue-shift factor is given by
\begin{equation}\label{uv}
\chi~\equiv~\frac{4G_NM}{\rho}~,
\end{equation}
where $\rho$ is the distance to the horizon.
Hence
\begin{equation}
\beta_{\text{local}}~=~\beta_{\text{Rindler}}\chi^{-1},~~\beta_{\text{local}}~\to~0,~~\text{if}~~\rho~\to~0.
\end{equation}
Note that if we go beyond the
Rindler approximation the overall Euclidean thermal manifold is {cigar-shaped}.

\subsection{Black holes and limiting temperature}
In quantum field theory the entropy density  $s~\sim~T^3$
and by using (\ref{uv}) the total entropy stored near the horizon is estimated as 
\begin{equation}
S~\sim~\int d\rho T(\rho)^3~=~ \frac {\text{Area}}{\epsilon^2}~,
\end{equation}
where $\epsilon$ is a cut off at small distances.

Not to exceed the black hole entropy (\ref{bhentropy}), we
 need a limiting temperature (brick wall of 't Hooft \cite{thooft}\cite{'tHooft:1984re}).
In other words, there is a need for a modification of  QFT at short distances.
Strings are welcome back on the fundamental level!

In practice, to limit applicability of field theory near the horizon
one introduces a so-called 
stretched horizon. The stretched horizon 
 is a surface placed close to the actual horizon, in front of it, such that
$G_{00}~\ll~1$.  For more details see \cite{damour}, \cite{parikh}. 
 
\section{Stringy horizon}
\subsection{Long-string picture of L. Susskind}

Consider the formation of a black hole by throwing in matter focused inside
the (future) black hole.
For a distant observer, the matter falls in infinitely long.
As a result,
  the infalling matter spreads out
in the transverse directions \cite{susskind}.
Indeed, consider the parton-model representation of the matter.
Then there is diffusion of the partons in the transverse directions.
And since the process takes long, the partons cover the whole area.
Moreover, it is known that, say, two long strings merge into a single one,
because of entropic considerations.
In this way one comes to the long-string picture of L. Susskind.
According to this picture near the horizon, at $\rho\sim l_s$ where $l_s$ is the
string scale, there exists a single long string \footnote{It is worth emphasizing that
physical pictures in general relativity are observer-dependent. The infalling matter is visualized
as a single long string by a distant observer.}.

Moreover, one might hope that by counting the number of states of
the long string, one could reproduce the entropy
(\ref{bhentropy}) of the black hole:  
\begin{equation}\label{matching}
S_{\text{long string}}~=~S_{BH}~~~~~(?)
\end{equation} 
Note, though, that the matching (\ref{matching}) is not without problems.
Indeed, the density of states of a long string is exponential  in its
length, or mass $M$, see (\ref{exponential}). On the other hand, the Bekenstein-Hawking
entropy is proportional to the area of the black hole, or its mass squared, 
$S_{BH}\sim M^2$. To maintain (\ref{matching}) one is forced to speculate that
this apparent mismatch is removed by accounting
 for the self-gravitation of the long string \cite{polchinski}. 

Thus, the long-string picture has definite advantages, by resolving 
the ultraviolet divergence  (\ref{uv}) through the introduction of a finite $l_s$
and by relating the number of degrees of freedom of a black hole to
the number of degrees of freedom of a long string
(which is much better understood). Also, the proportionality of the entropy to the
area comes out naturally because of the random walk of partons in the
transverse directions.
 
Many questions are left open, however.
In particular:
\begin{itemize}
\item{What keeps  the long string at $\rho~\sim~l_s$?}
\item{How to get quantitatively $S~=~(\text{Area})/4G_N$?}
\item{Qualitative picture vs fundamental strings?}
\end{itemize}
These questions were addressed in  \cite{mertens}, see
also references therein.

\subsection{Main results}

We considered a kind of mean field approximation, when a thin shell of matter 
of mass $\delta M$ falls into the black hole of mass $M$. The gravitational field of the mass $M$
is taken into account while the self-interaction within the shell is neglected.
The thin shell is described in the framework of string theory. 
We find that in Euclidean time the shell occupies a zero mode. 
This allows us to evaluate the entropy carried to the black hole by the shell.
Integrating over $dM$ reproduces the Bekenstein-Hawking entropy for black holes.
Moreover, knowing the wave function of the zero mode allows us to visualize the
profile of the stringy horizon at distances of order $l_s$ from the horizon.
One can summarize the results by saying that, in the approximation of the mean field,
the idea of relating the entropy of (large) black holes to the stringy degrees of freedom
is realized on a fully quantitative level. An important reservation is that these results
hold in the case of type II superstrings and heterotic strings in Rindler space.
Whereas for bosonic strings, there arises no consistent picture.

\subsection{Main tool: thermal scalar}

On the technical side, the results just mentioned are obtained
by utilizing the construction of the so-called thermal scalar in curved space,
see \cite{mertens1} and references therein. 
The theory of the thermal scalar has been developing since long, see
in particular \cite{kogan}, \cite{atick}.
Roughly speaking, the thermal scalar is a Euclidean counterpart
of the Hagedorn transition. Namely,
the original mechanism for the Hagedorn divergence of 
the partition function (\ref{divergence}) is the production
of high-mass states. 
A complementary view on the Hagedorn transition
in Euclidean time is that it is a kind of a Higgs phase transition
when the mass squared of a complex field changes its sign.
This scalar field, or thermal scalar, lives in spatial dimensions only (not temporal)
and its mass is given by: 
\begin{equation}\label{thermalscalar}
m^2_{\text{thermal scalar}}~=
~\frac{\big(\beta-\beta_{H}\big)
\beta_{H}}{2\pi(\alpha')^2}.
\end{equation}
Eq. (\ref{thermalscalar}) holds in flat space and as far as $m^2_{\text{thermal scalar}}$ is positive.
What happens at $\beta<\beta_H$ is not clear a priori, for further discussion see \cite{atick}.

The equivalence between the standard formulation of the Hagedorn transition
and that one in terms of the thermal scalar can readily
be demonstrated using the {polymer, or random-walk}
formulation of Euclidean field theory.
It is also straightforward to see that the thermal scalar corresponds to 
a string once wrapped
around the compact Euclidean time \cite{kogan}. The time dependence is fixed then by the
periodicity and effectively the wave function of the
thermal scalar depends only on the
spatial coordinates.

To consider strings in the background field of black holes, one needs 
to generalize the thermal scalar to curved space, 
see \cite{lawrence}, \cite{mertens1} and references therein.

\subsection{Thermal scalar in curved space}
The 
action for the thermal-scalar field $\varphi$ is given by
\cite{giveon,mertens1}:
\begin{eqnarray}\label{action}
S~=~\int d^{D-1}x\sqrt{G}e^{-2\Phi}~~~~~~~~~~~~~~~~~ \nonumber \\
\times\Big(G_{ij}\nabla^i\varphi\nabla^j\varphi^{*}+ 
\frac{1}{4\pi^2(\alpha')^2}(\beta^2G_{00}-\beta_{H}^2)\varphi\varphi^{*}\Big)~,
\end{eqnarray}
where $G_{00}, G_{ij}$ are the components of the metric and $\Phi$ is the dilaton field.
A crucial point is that for type II strings in Rindler space, $ds^2_{\text{Rindler}}= a^2\rho^2dt^2+d\rho^2+d{\bf x}_{\perp}^2$,
the action (\ref{action}) receives no $\alpha'$
corrections \cite{giveon,mertens1}. Moreover,  this is also true for
heterotic strings \cite{mertens2} (but not for bosonic strings).

The corresponding equation for the wave functions
\begin{equation}
\Big(-\partial_{\rho}^2-\frac{1}{\rho}\partial_{\rho}+
\frac{1}{4\pi^2(\alpha')^2}(\beta^2a^2\rho^2-\beta_{H}^2)\Big)
\varphi_n(\rho)~=~\lambda_n\varphi(\rho)
\end{equation}
has solutions:
\begin{equation}\label{wf}
\varphi_n(\rho)~=~\exp\Big(-\frac{a\beta\rho^2}{4\pi\alpha'}\Big)
L_n\Big(\frac{a\beta\rho^2}{2\pi\alpha'}\Big), 
~\lambda_n~=~a\beta(1+2n)-2\pi
\end{equation}
in terms of Laguerre polynomials $L_n$, and $n\in \mathbb{N}$.
The zero mode ($n$=0) at $\beta=\beta_{\text{Rindler}}=2\pi/a$
dominates the thermal partition function.

This analysis was done for the singly wound thermal scalar state. A curiosity at this point is that states that are wound multiple times are simply absent from the spectrum. This will be important in the final section.

\subsection{Picture emerging}
The build-up of a black hole by throwing a thin shell of mass $\delta M$ to the black 
hole of mass $M_{\text{initial}}$
can be consistently described by string theory in
a mean-field 
approximation. The shell ends up as a 
long string in a layer of thickness $\delta\rho~\sim l_s$
near the horizon.

 Moreover, the entropy of black holes is calculable
without any adjustable parameters. In more detail, the
 density of states seen by a distant observer is
\begin{equation}
\omega(\delta M)~\sim~\frac{\exp (\beta_{\text{Hawking}}\delta M)}{\delta M},~\text{or}~
\beta_{\text{Hawking}}=\beta_{\text{Hagedorn}}.
\end{equation} 
Integrating over $\delta M$ we get the Bekenstein-Hawking entropy:
\begin{equation}
\delta S_{BH}~=~8\pi G_NM\delta M~\to~S_{BH}~=~\frac{\text{Area}}{4G_N}.
\end{equation}
Also, knowledge of the wave function (\ref{wf}) allows us
to evaluate \cite{mertens4,mertens} the profile 
of the energy-momentum tensor associated with
the long string of mass $\delta M$:
\begin{align}\label{profile}\nonumber
\left\langle T^0_0({\bf x})\right\rangle &=~-2N^2\Big(\frac{2\rho^2}{\alpha'}-1\Big)e^{-\rho^2/\alpha'}~,\\ 
\left\langle T^{\rho}_{\rho}({\bf x})\right\rangle &=~2N^2e^{-\rho^2/\alpha'}~,\\\nonumber
\left\langle T^{i}_{j}({\bf x})\right\rangle &=~2N^2\delta^i_j
\Big(1-\frac{\rho^2}{\alpha'}\Big)e^{-\rho^2/\alpha'}~,
\end{align}
where $i,j=1,2$ are transverse directions and
 $N$ is a normalization factor.

Note a positive value of the radial pressure, $\left\langle T^{\rho}_{\rho}({\bf x})\right\rangle$.
It is this pressure that keeps the matter from collapsing onto the center.
Remarkably, Eqs (\ref{profile}) specify a distribution of matter near
the horizon, at $\rho\sim l_s$.

\section{From strings to gauge theories, via holography}

Amusingly enough, lessons from strings on the gravitational scale might 
produce a new insight into the dynamics of gauge theories.
The means is holography: strings live in curved extra dimensions, while gauge
theory lives on a flat boundary. Both theories are inter-related.

The most famous case of the string-gauge duality refers to  
$\mathcal{N}=4$ supersymmetric Yang-Mills theory \cite{maldacena}.
In case of non-supersymmetric Yang-Mills theories, Witten constructed 
a model \cite{witten} which in the far infrared limit belongs to the same
universality class as large-$N_c$ gauge theories.
The model can be generalized to incorporate massless quarks  
\cite{sakai}.
From our perspective, it is crucial  that the geometry  
in extra dimensions inherent to this model is of the same cigar-shape 
as we encountered while discussing large black holes above.

In more detail, the Euclidean version of the model \cite{witten}
introduces
 {two compact dimensions}:
\begin{itemize}
\item{Periodic Euclidean time $\tau~\sim~\tau~+\beta_{\tau}(z)$ where the
periodicity, $\beta_{\tau}$ depends on an extra coordinate $z$. As 
is common  to holographic models, the $z$-coordinate 
is associated with resolution. The limit $z\to 0$ 
corresponds to Yang-Mills theories in the ultraviolet limit, while 
$z\to~z_{\text{horizon}}$ corresponds to the infrared limit on the field theoretic
side}
\item{There is another periodic coordinate $\sigma~\sim~\sigma+\beta_{\sigma}(z)$.
Wrapping around $\sigma$ {counts the topological charge} associated with the
corresponding
stringy state}
\end{itemize}

From first principles, at  $T=0$ the $(\tau, z)$ space is a cylinder
and $(\sigma, z)$ is cigar-shaped, $\beta_{\sigma}(z_{\text{horizon}})=0$.
At  the deconfining phase transition, $T=T_{cr}$ the geometries 
in the $(\tau, z)$ and $(\sigma, z)$ coordinates are interchanged.

Now we are coming to a central point, namely how, if at all, the geometry in
the extra dimensions is related to gauge theory phenomenology.
One of the routes is to consider properties of so-called defects.
One of the best studied examples of a defect on the field-theoretic
side are instantons. On the stringy side, defects can be identified topologically, 
in terms of wrapping around compact directions. In the geometric language
the string action is very simple, 
$$(\text{action})\sim L\cdot (\text{tension})~~.$$ 
If there is a cylinder-type geometry then the wrapping number is well-defined
and the probability to find a defect is suppressed by the action.

However, in case of a cigar-shaped geometry,
$$\beta_{\tau}(z_\text{horizon}) ~=~0~\text{or}~ \beta_{\sigma}(z_\text{horizon}) ~=~0$$
the action for wrapped states vanishes at the horizon
and the probability to find them in the vacuum state is not
suppressed by their action.  

Instantons are
distinguished by a non-vanishing topological charge. As is mentioned above,
in the geometric language, the topological charge is associated with wrapping around the $\sigma$-coordinate. The simplest geometric object
which can be wrapped around the $\sigma$ direction
is a $D0$-brane. The corresponding defect is characterized by a topological charge
determined by the wrapping number, its position in the Euclidean space and 
by its action, which depends on the value of its $z$-coordinate. Thus, the
number of the parameters characterizing such $D0$ branes matches
instantons of field theory, for further details and references see \cite{bergman}. 

Let us check the gravity-gauge correspondence 
on the example of the instantons \cite{bergman}. Consider first temperature $T=0$ and start
with the strings. Since $\beta_{\sigma}$
vanishes on the horizon, $\beta_{\sigma}(z=z_\text{horizon})=0$, instantons are not 
suppressed by the action in the far infrared. 
 This is, indeed, well known on the field theoretic side of the correspondence.
At $T = T_{cr}$ the $(\sigma, z)$ geometry is changed into a cylinder
and instantons become suppressed according to the stringy picture. 
Again, this conclusion is well known in field theory and is supported by the 
existing phenomenology.

 Now we are coming to the central point of this section,
 that is the manifestation of the zero mode discussed in the preceding
 section.  
Classically, the action for the
defects, $$S_{\text{defect}}~=~L\cdot (\text{tension})$$
vanishes for any wrapping number as far as $\beta_{\sigma}(z)=0$.
However, quantum-mechanically keeping a $D0$ brane at 
the tip of the cigar results in kinetic energy, because of the uncertainty principle.
As a result, only the lowest level survives as 
the zero mode at the tip of the cigar
(see discussion in the preceding section). 
The lowest level, in turn, corresponds to a single wrapping.
Phenomenologically,
this implies that only instantons with the topological charge
$Q_{\text{top}}=\pm 1$ exist while
$|Q_{\text{top}}|\geq 2$ are suppressed in the vacuum. Also, an effective
infrared cut off might arise dynamically in the far infrared.
These predictions which follow from holography seem to be supported by the lattice data.

\end{document}